\documentclass[aps,prb,showpacs,floatfix,amsmath,amssymb,
               reprint]{revtex4-1}
\usepackage{color}
\usepackage{graphicx}
\usepackage{natbib}
\usepackage{epsfig}
\usepackage{setspace}
\usepackage{amsmath}
\usepackage{amssymb}
\usepackage{verbatim}
\usepackage{bibentry}

\def\nnu{{\nonumber}}
\begin{document}
\title{Quantum critical Mott transitions in a bilayer 
Kondo insulator-metal model system}

\author{Sudeshna Sen}
\author{N.\ S.\ Vidhyadhiraja}\email{raja@jncasr.ac.in}
\affiliation{Jawaharlal Nehru Centre for Advanced Scientific Research,
Bangalore-560064, Karnataka, India}
\begin{abstract}
A bilayer system comprising a Kondo insulator coupled to a simple metal (KI-M) is considered.
Employing the framework of dynamical mean field theory, the model system is shown to exhibit a surface of  quantum critical points (QCPs), that separates a Kondo 
screened, Fermi liquid phase from a local moment, Mott insulating phase. The quantum critical nature of these
Mott transitions is characterized by the vanishing of (a) the
coherence scale on the Fermi liquid side, and (b) the Mott gap
on the MI side. In contrast to the usual `large to small' Fermi surface (FS) QCPs in heavy fermion systems, the bilayer KI-M system exhibits a complete FS destruction.
\end{abstract}

\maketitle

\section{Introduction}

The interaction driven Mott metal-insulator transition
(MIT)
is an integral aspect of a wide variety of phenomena exhibited by strongly 
correlated electron systems.
The generic scenario for Mott MITs is that they are
first order in nature, with a finite temperature critical
end point \cite{DMFT_RMP, Kotliar_free_energy, Park_CDMFT_2008, Gull_DCA}, 
and are usually accompanied by large hysteresis curves, as seen for example in  
experiments on   
the vanadium oxides, V$_2$O$_3$ and VO$_2$ \cite{VO2_2011,Limelette_expt}. 
The two possible exceptions are (a) a bilayer $\phantom{a}^3$He system~\cite{He3_unique_realizn_FL1}, 
that can be pushed to a 
layer selective quantum critical Mott transition by tuning the filling of the 
layers~\cite{He3_expt, Assad_2011_bilayerHM, Assad_2012_bilayerHM,
Pepin1, Pepin2, Pepin3}; and
(b) two-dimensional organic insulators, where a pressure
driven Mott quantum critical point (QCP) has been found very recently~\cite{Mott_QCP2015}.

The idea of a QCP in a  
Mott transition has thus far appeared as an exception rather than the rule, 
unlike in the case of heavy fermions that offer numerous examples
of quantum criticality, which are driven either by spin or valence fluctuations
 \cite{Vojta_review_QCP, Si_Steglich_QCP_review}. 
A recent work by Senthil \cite{Senthil1} proposed the possibility of 
observing a second order 
Mott transition from a Fermi liquid metal to a Mott insulator with a spinon 
Fermi surface in two dimensions. 
The role of critical fermionic quasiparticles has 
been proposed in the context of continuous Mott transitions and orbital 
selective Mott transitions at zero temperature 
\cite{Senthil_Vojta_QCP1, Senthil1, Vojta_review_QCP}. 
Recent study of quantum critical transport near the Mott transition in 
half-filled Hubbard model \cite{Hanna_Mott_criticality} has been 
interpreted as evidence of a hidden Mott quantum criticality.
Further study on doped Hubbard models reveal a clear connection between 
bad metal phenomenology and signatures of Mott quantum criticality.
\cite{DMFT_Mott_QCP2015}. Studies of 
holographic duality propose the existence of yet-unspecified  
quantum critical regimes for strange metals in cuprates 
\cite{Sachdev1, *Sachdev2}. 
In order to obtain a 
better understanding of the continuous Mott MIT, it would be desirable
to identify new systems where Mott quantum criticality may be observed.

In this paper, we consider a  bilayer heavy fermion system comprising a 
Kondo insulator (KI), represented by a symmetric periodic Anderson model (PAM) 
\cite{DMFT_RMP, Raja_dyn_sca_pam}, 
coupled
to a non-interacting metal (M).
Previous theoretical studies on coupling a KI layer to metallic layers have 
shown\cite{SSen2015, Peters1} that the
normally gapped $f$ and $c$-bands of the KI acquire a finite density of states
and a quadratically vanishing gap respectively due to the coupling. 
This emergent metallic phase of the $f$-electrons
leads us to anticipate an interaction driven Mott transition. To our surprise, 
we find a plethora of interaction driven Mott QCPs in the KI-M system. 
The quantum critical nature of the Mott transitions is
established by the vanishing of the coherence scale from the Fermi
liquid side, and of the Mott gap from the insulating side at the
same point in the parameter space.
The competition between itinerancy and localization
might lead to as yet unknown universality classes for the Mott
transition \cite{Sachdev1, *Sachdev2}, and indeed, the system
under consideration, might serve as an ideal model to investigate
the universality and scaling associated with Mott quantum criticality.

\section{Model, Theoretical Framework and Analytical results}

The Hamiltonian of the bilayer KI-M model system is given by,
\begin{align}
 \mathcal{H}
 &= \sum_{\mathbf{k}i\sigma}\begin{pmatrix} 
 f^\dagger_{i\sigma} & c^\dagger_{\mathbf{k}\sigma} & c^\dagger_{M,\mathbf{k}\sigma}
\end{pmatrix}
\begin{pmatrix} 
\epsilon_f & 0 & 0 \\
0 & \epsilon_c+\epsilon_\mathbf{k} & 0 \\
0 & 0 & \epsilon^{\phantom\dagger}_{cM} + \epsilon_\mathbf{k}
\end{pmatrix}
\begin{pmatrix} 
 f^{\phantom\dagger}_{i\sigma} \\ c^{\phantom\dagger}_{\mathbf{k}\sigma} \\ c^{\phantom\dagger}_{M,\mathbf{k}\sigma}
\end{pmatrix} \nonumber \\
& + U \sum_i n_{fi\uparrow}n_{fi\downarrow}+ 
V \sum_{i\sigma}\left(f^{\dagger}_{i\sigma}c_{i\sigma}+\mathrm{H.c.}\right)\nonumber \\
& + t_\perp\sum_{i\sigma}\left(c^\dagger_{M,i\sigma}c_{i\sigma}+\mathrm{H.c.}\right)
\label{eq:ham}
\end{align}
where $\epsilon_f$ is the orbital energy of the non-dispersive $f$-band, 
which is coupled to a conduction band with dispersion $\epsilon_\mathbf{k}$ and orbital 
energy $\epsilon_c$ through a hybridization $V$. These $f$-and $c$-orbitals 
belong to the KI layer, which is coupled to the metallic layer, characterised 
by a conduction band, $\epsilon_{\mathbf{k}}$, and a orbital energy,
$\epsilon_{cM}$, through an inter-layer hopping, $t_\perp$ between the 
$c$-orbitals of the layers. 
The second term,  represents the cost
of double occupancy, $U$, of the $f$-orbitals of the KI layer. 

In this work, we consider the particle-hole limit of this model where
$\epsilon_c=\epsilon_{cM}=\epsilon_f + U/2=0$, and each of the three bands
are half-filled.
The KI-M model may be solved exactly in the non-interacting limit ($U=0$),
and the atomic limit ($V=0$)~\cite{SSen2015}. In order to investigate
the system beyond these simple limits, we employ the DMFT 
framework \cite{DMFT_RMP},
within which the self-energy becomes local or
momentum-independent. 
  A cavity construction \cite{DMFT_RMP} for the bilayer model considered in 
Eq.~\eqref{eq:ham} yields the following effective action for the 
$f$ electrons:
  \begin{align}
    \label{eq:action}
    S_{eff}=&-\int_0^\beta d\tau\int_0^\beta d\tau'\sum_\sigma
    f_\sigma^{*}(\tau)\mathcal{G}_{0f}^{-1}(\tau-\tau')f_\sigma(\tau)+\nnu\\
    &U\int_0^\beta d\tau n_{f\uparrow}(\tau) n_{f\downarrow}(\tau), \\
 {\rm with,} & \nnu \\
    \label{eq:Gf0}
    \mathcal{G}_{0f}(\omega)=&\int\frac{\rho_0(\epsilon)}
    {\omega^{+}-\frac{V^2}{\omega^+-\epsilon-t_\perp^2/(\omega^+-\epsilon)}}
     d\epsilon
\end{align} 
The $f$-Green's function may be written as
$G^f(\omega) = \sum_k G^f(\omega,\epsilon_k)=\int_{-\infty}^{\infty} 
d\epsilon\,\rho_0(\epsilon) G^f(\omega,\epsilon)\,$, 
where, the $\rho_0(\epsilon)=2\sqrt{1-\epsilon^2/D^2}/(\pi D)$ is a 
semi-elliptic non-interacting conduction band density of states (DoS) 
of width 
$2D$, appropriate for a system with long-range
hoppings, chosen to suppress magnetic ordering \cite{DMFT_RMP}; 
\begin{align}
G^f(\omega,\epsilon)=\frac{\rho_0(\epsilon)}
    {\omega^{+}-\frac{V^2}{\omega^+-\epsilon-t_\perp^2/(\omega^+-\epsilon)}
-\Sigma_f(\omega)}
\label{eq:Gf}
\end{align}
with $\Sigma_f(\omega)$ being the Hartree corrected, momentum-independent $f$-self-energy.
 Several results may be derived from
these expressions, that are outlined below. 
For $U=V=0$, the bilayer system undergoes a metal to
band-insulator transition when the interlayer hopping 
exceeds the conduction band width.
However, 
with $V\ne 0$, the non-interacting $f$-density of 
states, given by $D^f(\omega; U=0)=-{\rm Im}G^f(\omega;U=0)/\pi$
 does not vanish for any $t_\perp\neq 0$, although
 the $\omega=0$ peak in the density of states gets progressively narrower with increasing $t_\perp$:
$D^f(\omega; U=0) 
\stackrel{t_\perp\gg 1}{\longrightarrow}  
\frac{t_\perp^2}{V^2} \rho_0\left(\frac{\omega t_\perp^2}{V^2}\right).$
From the above expression, it is easy to see that, when $t_\perp\gg 1$,
the effective bandwidth, $D_{eff}$ of the $f$-DoS  
is greatly reduced as $\sim V^2/t_\perp^2$, hence 
the interaction strength needed for the Mott metal to
insulator transition may be estimated as 
\begin{equation}
U_c \sim D_{eff} \stackrel{t_\perp \gg 1}{\longrightarrow}
D\,V^2/t_\perp^2,
\label{eq:asymptote}
\end{equation} 
implying that the interaction strength
required to drive the Mott transition should decrease
 rapidly with increasing inter-layer hopping. 

The $c$-electrons in the heavy fermion 
layer acquire a self-energy, $\Sigma_{c1}(\omega)$ that can be related to 
$\Sigma_f(\omega)$ as, 
$\Sigma_{c1}(\omega)=\frac{V^2}{\omega^+-\Sigma_f(\omega)}$. 
For, the non-interacting metallic layer, $V=0$ and hence,  
$\Sigma_{c2}(\omega)=0$.
In our previous work~\cite{SSen2015} we had shown that for a bilayer system, 
the $c$-electron Greens functions are given by,
\begin{align}
G_{11}(\omega)&= a\,H[\Gamma_+] +b\,H[\Gamma_-],  \label{eq:bilayer1}\\
G_{22}(\omega)& = a^\prime\, H[\Gamma_+] + b^\prime\, H[\Gamma_-],\label{eq:bilayer2}
\end{align}
where
\begin{align}
\Gamma_\pm &= \frac{1}{2} \left(\lambda_1 + \lambda_2 \pm 
\sqrt{(\lambda_1-\lambda_2)^2 + 4t_\perp^2}\right),
\label{eq:gamma_pm}\\ 
\lambda_r &=\omega^+ - \epsilon_c - \Sigma_{cr}(\omega),\;\;r=1,2\nnu
\end{align}
\begin{align}
a &= \frac{\Gamma_+ - \lambda_2}{\Gamma_+ - \Gamma_-}
\;\;{\rm and}\;\;b = \frac{\lambda_2 - \Gamma_-}{\Gamma_+ - \Gamma_-}, \\
a^\prime &= -\frac{\lambda_1 - \Gamma_+}{\Gamma_+ - \Gamma_-}
\;\;{\rm and}\;\;b^\prime = \frac{\lambda_1 - \Gamma_-}{\Gamma_+ - \Gamma_-},
\label{eq:coeff}
\end{align}
and $H[z]$ represents the Hilbert transform with respect to the
bare DoS of the conduction electrons. 
Using these expressions one may then derive a low energy form of the 
respective spectral function. In the low frequency regime of a
Fermi liquid (FL) 
phase, where the $f$ self-energy may be expanded in a Taylor series as:
$\Sigma_f(\omega)=\Sigma_f(0) + \omega (1-1/Z) + {\cal{O}}(\omega^2)$ 
(where $Z$ is the quasiparticle weight),
the DoS of the $c$-e$^-$s of the heavy fermion layer has the following form:
\begin{align}
  D^{c}_1(\omega)& \stackrel{\omega\rightarrow 0}{\sim} 
  \left(\frac{\omega t_\perp}{ZV^2}\right)^2  
  \rho_0\left(\omega\left[1+\frac{t_\perp^2}{ZV^2}\right]\right) \label{eq:Dc} \\ \nonumber
  & + \left(1 - \left(\frac{\omega t_\perp}{ZV^2}\right)^2\right) 
  \rho_0\left(\omega\left[1-\frac{t_\perp^2}{ZV^2}\right] 
  - \frac{ZV^2}{\omega}\right)\,.
 \end{align}
Further, since  $D^f(\omega)=(Z^2V^2/\omega^2) D_{c}(\omega)$, 
the $f$-e$^{-}$ DoS at $\omega=0$ is
pinned to the non-interacting 
limit namely, $V^2D^f(0)/t_\perp^2 = \rho_0(0)$, thus exhibiting 
adiabatic continuity to the non-interacting limit and hence 
satisfying a necessary condition for a Fermi liquid (FL).
In a Mott insulating phase, on the other hand,
the self-energy  must have a pole at $\omega=0$ with the
form $\Sigma_f(\omega) \stackrel{\omega\rightarrow 0}{\longrightarrow} 
\alpha/\omega^+$,
where the residue of the pole is 
$\alpha \sim {\cal{O}}(U^2)$ \cite{DMFT_RMP, hbar_raja2011}. 
So, if the bilayer system should have a Mott insulating phase, the 
denominator of $G^f(\omega;\epsilon)$
should not have zeroes for any $|\epsilon| < D$ as 
$\omega \rightarrow 0$, since that would imply a non-zero density of states in 
the single-particle spectrum at the chemical potential. It may be shown easily 
that such a condition is satisfied only for $t_\perp > D$. This condition can 
also be derived by substituting, 
$\Sigma_f(\omega)\stackrel{\omega\rightarrow 0}
{\longrightarrow}\alpha/\omega^+$, in Equation~\eqref{eq:bilayer2}.
Thus, a Mott insulating phase can be sustained only if $t_\perp>D$.

From the above analysis, we can anticipate a line of Mott transitions in 
the  $t_\perp-U$ plane, where the critical $U$ should decrease at least as 
rapidly as $1/t_\perp^2$ for $t_\perp \gg 1$ and for $U\rightarrow \infty$, 
the critical $t_\perp$ should asymptotically approach the bandwidth $D$. 
In order to determine the phase diagram quantitatively,
we must obtain the self-energy $\Sigma_f(\omega)$ and examine the 
quasiparticle weight as well as the Mott gaps. In this work, we choose the 
local moment approach (LMA) as the solver for the $T=0$ 
quantum impurity problem. It is a diagrammatic perturbation theory 
based approach, which is built around the two broken-symmetry, 
local moment solutions ($\mu=\pm |\mu_0|$) of 
an unrestricted Hartree-Fock mean field approximation. The spin-flip dynamics 
 are subsequently incorporated through an 
infinite order resummation of a specific class of diagrams that represent the 
spin flip process necessary to capture the Kondo effect. 
Finally, a key ingredient of LMA is to impose adiabatic continuity 
to the noninteracting limit, that is crucial for the recovery of Fermi 
liquid behaviour and the emergence of a low energy scale. This is known as 
the symmetry restoration condition and any violation of this condition is a 
signal of a quantum phase transition to another phase such as a local moment 
phase.
More details of the LMA may be found in previous work\cite{LMA_SIAM1, *LMA_KIs,Raja_dyn_sca_pam}. 
The LMA has been shown to benchmark excellently 
against numerical renormalization group (NRG) \cite{LMA_NRG_benchmark1}, 
and Bethe Ansatz for the single impurity Anderson model and the Kondo problem respectively 
\cite{LMA_SIAM1}. Subsequently, it has been employed in studies on Kondo insulators 
and heavy fermion systems, where the dynamics and transport properties of several heavy 
fermions systems were quantitatively explained\cite{*LMA_KIs,Raja_dyn_sca_pam, 
Raja2_2005}. 
The LMA has also been used in studying 
specific cases of impurity systems with many orbitals \cite{LMA_SU2_AIM}, 
the pseudogap Anderson model, the gapped 
Anderson impurity model and the soft-gap Anderson model 
\cite{LMA_pseudogap_AIM1, *LMA_pseudogap_AIM2, LMA_softgap, LMA_GAIM}. 
 The study on the soft gap Anderson model has also been benchmarked with NRG 
\cite{LMA_NRG_soft_gap_AIM}. The resultant LMA phase boundary was in good 
quantitative agreement with NRG results \cite{LMA_GAIM}.
However, generalizing LMA for employing it as a cluster
solver to incorporate non-local dynamical correlations
has not been attempted yet.

\begin{figure}[h!]
   \centerline{\includegraphics[clip=,scale=0.6]
                        {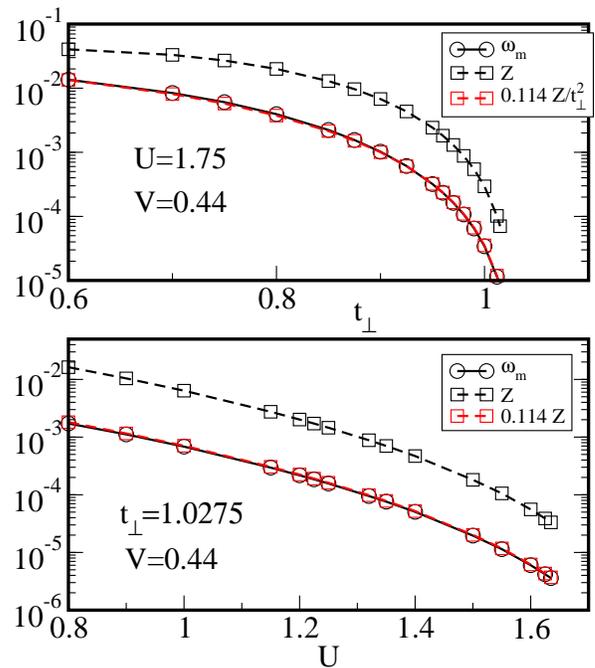}}
   \caption{{\bf Comparison of the low energy spin flip scale, $\omega_m$ 
   and the quasi-particle weight, $Z$:} $\omega_m$ (black circles) 
            and the corresponding quasiparticle weight, $Z$ (black squares), 
            shown on the same scale. A simple multiplicative rescaling of the 
            $Z$ makes it identical to the $\omega_m$ data as shown by the red 
            squares. } 
   \label{fig:wm_Z}
 \end{figure}
  \noindent
\section{Numerical Results}

\subsection{Phase diagram}
 
  As discussed earlier, the LMA captures the Kondo effect through the 
  calculation of the transverse spin-flip processes embodied in the 
  infinite-order resummation of a specific class of 
  diagrams \cite{LMA_SIAM1}.  
  The low energy spin flip scale, `$\omega_m$', thereby generated and
  identified through the 
  position of the peak of the imaginary part of the transverse
  spin polarization propagator within LMA, is known to be proportional to 
the Kondo scale in the single impurity Anderson model \cite{LMA_SIAM1}. 
  The `$\omega_m$' scale is 
  proportional to the quasiparticle weight ($Z=\left[1-\frac{\partial \mathrm{Re}\Sigma(\omega)}{\partial\omega}\right]^{-1}$) also, as shown in 
Fig.~(\ref{fig:wm_Z}), 
  where both 
  $Z$ and the 
  $\omega_m$ are plotted and rescaled with simple multiplicative factors to highlight their  equivalence.
\begin{figure}[t]
   \centerline{
   \includegraphics[clip=,scale=0.55]
                        {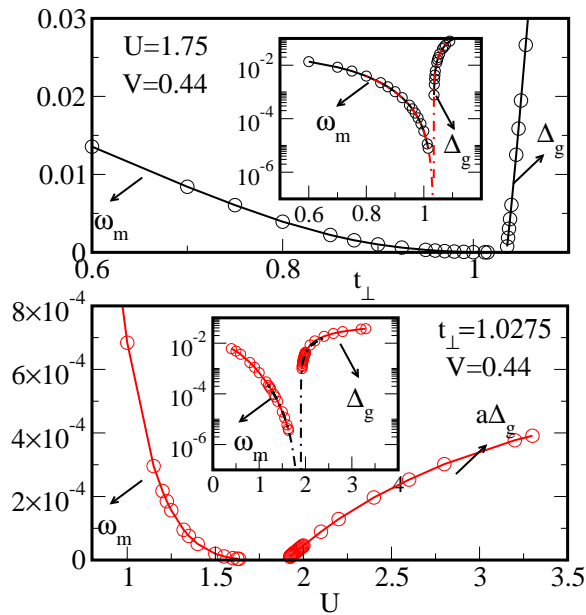}}
    \caption{\textbf{
             The low energy spin flip scale, $\omega_m$ and the Mott gap, 
$\Delta_g$ for $V=0.44$ as a function of $t_\perp$ (top panel, $U=1.75$, 
$t_{\perp c}\approx1.03$) and  
$U$ (bottom panel, $t_\perp=1.0275$, $U_c\approx1.9$).} 
Note that in the main part of the bottom panel, the Mott gap values 
have been scaled by a multiplicative factor ($a=1/95$) to show the $\omega_m$ 
and $\Delta_g$ on the same scale. In the insets, the same is plotted on a 
linear-log scale but the Mott gap is not rescaled. The extrapolation (dot-dashed line) to the 
critical value 
is shown to highlight that the Kondo scale and the Mott gap 
approach zero at the same critical point.}
    \label{fig:wm_gap}
   \end{figure}

 Thus, we choose $\omega_m$ to be the coherence scale in the Fermi liquid 
 phase. The vanishing of the $\omega_m$ scale then signifies the divergence of 
 the effective mass at the Mott transition. In the Mott insulating phase, this 
 low energy scale vanishes, i.e.\  $\omega_m=0$ indicating a zero energy cost 
 to flip an unscreened local moment and a Mott gap develops.
 The Mott gap then serves as the relevant 
 energy scale in the Mott insulating phase. Conventionally, one then 
 probes the behaviour of these two relevant energy scales as a function of 
 $U$ to determine if there exists any regime of $U$ where the metallic and the 
 Mott insulating states may coexist, which would imply a first order Mott 
 transition \cite{preformed_gap1, preformed_gap2, DMFT_RMP, Limelette_expt, Mott_MIT_exact}. 

 \begin{figure*}[t!]
  \centerline{\includegraphics[clip=,scale=0.325]
                        {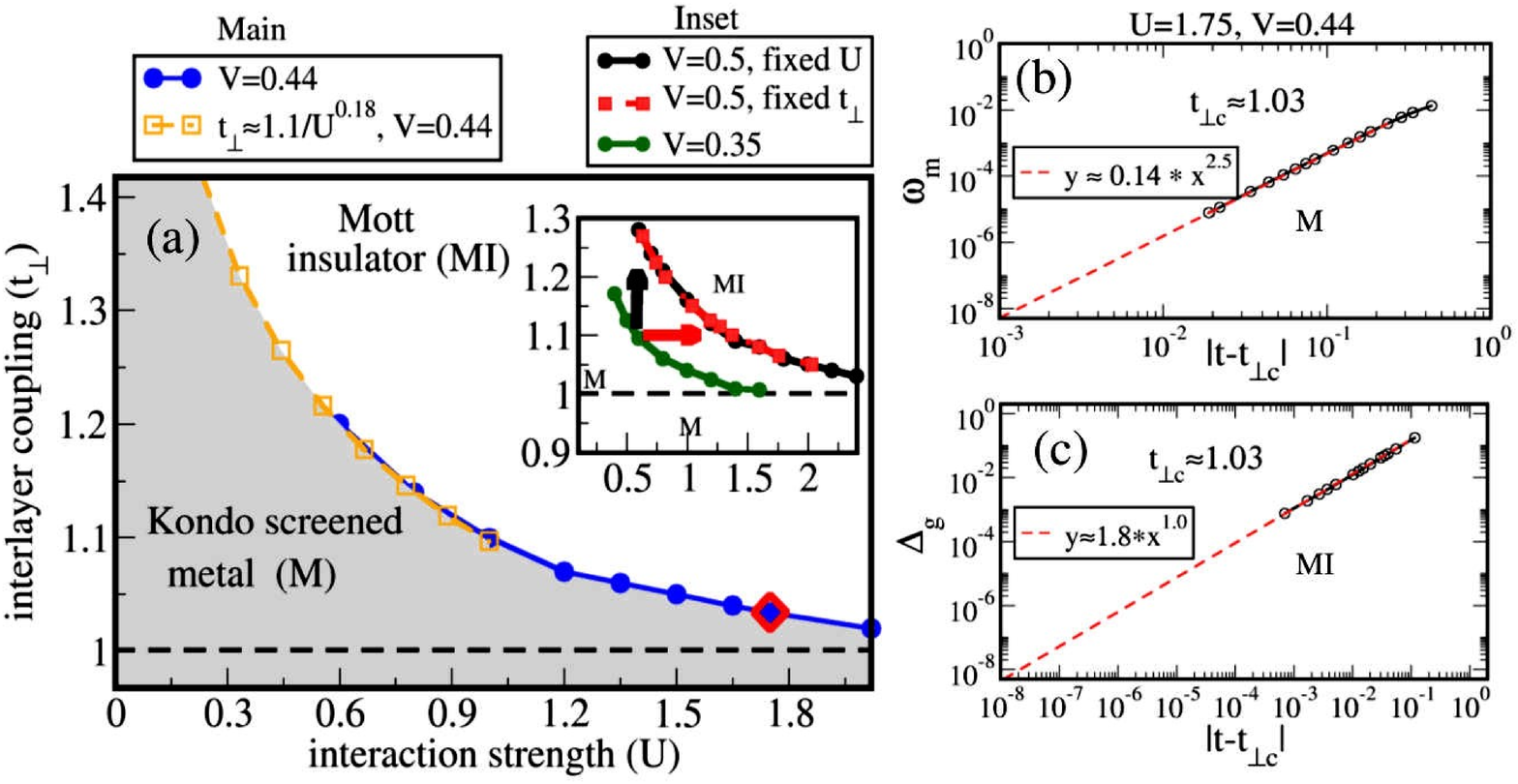}}
   \caption{\textbf{The phase diagram in the $t_{\perp}-U$ plane for 
                    the transition from a Kondo screened Fermi liquid 
                    to a Mott insulating phase}: 
            (a) The main panel shows the line of QCPs obtained, for $V=0.44$ 
(blue circles), and, the inset shows the same for $V=0.5$ 
(black circles, red squares) and $V=0.35$ (green circles).  
  While for $t_\perp\gg1$ it was estimated 
  that $t_{\perp c}\sim \sqrt{DV^2/U_c}$, the numerical data, corresponding to 
low $U$ asymptote, fits well to a form $t_{\perp c}\sim f(V)/U_c^{0.2}$ as 
represented by the
  orange dashed line with squares for $V=0.44$ (also see main text). 
The black dashed line 
  represents the $t_\perp=D$ boundary for the existence of the Mott 
  transition. 
Additionally, in the inset, we also show, (for $V=0.5$) that the line of 
quantum critical Mott transitions 
is the same irrespective of the direction of approach, namely, $t_\perp$ 
driven (black circles, black arrow) or $U$ driven (red squares, red arrow).
(b, c) In order to identify the $t_\perp c$ or the $U_c$, we plot the 
$\omega_m$ (shown in (b) for $t_\perp c$) or the $\Delta_g$ 
(shown in (c) for $t_{\perp c}$) on a log-log scale. 
Starting from an initial guess value, $t_{\perp c}^{guess}$, 
we then add/subtract a small number $\sim0.001$ from the simulated data points, 
plotted on a log-log scale. This subtraction/addition is continued until 
a visually inspected straight line was obtained that was fitted to the form 
$y=a|x-x_c|^{\alpha}$. The same procedure is repeated for $U$ driven 
transitions. The representative plot corresponds to $U=1.75,\; V=0.44$ that 
also represents the top panel of Fig.~(\ref{fig:wm_gap}). The corresponding 
extrapolated point is also highlighted as a red diamond in (a). 
The same is repeated for several other parameters and the phase diagram (a) 
in the $t_\perp$-$U$ plane is generated.}
   \label{fig:phase_diag}
 \end{figure*}
 
 In Fig.~(\ref{fig:wm_gap}) we therefore show the behavior 
  of the FL scale $\omega_m$, in the Kondo screened phase and the Mott gap, 
  $\Delta_g$ 
  in the Mott insulating state as a function of
  $t_\perp$ (top panel) and as a function of $U$ (bottom panel). 
  Indeed, we find that the two scales vanish as power laws given by 
  $\omega_{m}\sim (g_c-g)^{\alpha_{\scriptscriptstyle{FL}}}$ and 
  $\Delta_g\sim (g-g_c)^{\alpha_{MI}}$, where the athermal parameter, $g$, 
  represents $t_\perp$ or $U$. 
  It is seen from the top and bottom panels that either of the scales vanish
  at precisely the same (extrapolated) $t_{\perp c}$ (top) or $U_c$ (bottom), 
  thus indicating a quantum critical point in the $t_\perp - U$ plane. 

  A similar analysis for a wide range of values of $U, V$ and $t_\perp$ yields 
  a family of lines of QCPs in the $t_\perp-U$ plane for various $V$ values, 
  as shown in Fig.~\ref{fig:phase_diag}(a). In  Fig.~\ref{fig:phase_diag}(a) 
we show the line separating a Kondo screened phase from the Mott insulating 
phase, for different $V$'s, namely, $V=0.44$ in the main panel and 
$V=0.35,\;0.5$ in the inset. These lines of QCPs represent a $t_\perp$ driven 
Mott transition and the procedure adopted to obtain the $t_{\perp c}$ at a 
fixed $U$ is demonstrated in Fig.~\ref{fig:phase_diag}(b, c). 
Additionally, 
we also demonstrate the line of QCPs corresponding to a $U$ driven transition 
for $V=0.5$ as represented as red squares in the inset of 
Fig.~\ref{fig:phase_diag}(a), found to be identical with the critical points 
obtained for a $t_\perp$ driven transition. 
This confirms that $U$ and $t_\perp$ would drive 
the respective Mott MITs at the same critical point, irrespective of the 
direction of approach. In Fig.~\ref{fig:phase_diag}(a) (main panel), we find that 
while the curve {\it qualitatively} represent the analytical insight due to 
Eq.~\eqref{eq:asymptote};  however, the high $t_\perp$ 
and low $U$ asymptote appears to fit well with a form, 
$t_{\perp c}\sim f(V)/U_c^{0.2}$. While this quantitative discrepancy could 
stem from the approximate analytical argument, it could also be possible 
that we need to go to even lower $U$'s and even higher $t_\perp$'s to 
extract the form of the asymptote obtained from the numerical data.

We now mention the procedure adopted to extract the 
critical points in detail. For 
demonstration, we use the same data as shown in the top panel of 
Fig.~(\ref{fig:wm_gap}). In the Fermi liquid 
phase (Fig.~(\ref{fig:phase_diag})(b)), 
the calculations were performed until a Kondo scale of 
$\sim 5\times10^{-5}$ 
was reached. Starting from an initial guess value,  
$t_{\perp c}^{guess}$ of the respective athermal variable, we then 
added/subtracted a small number $\sim0.001$ from the simulated data points, 
plotted on a log-log scale. This subtraction/addition was then continued until 
a visually inspected straight line was obtained on the log-log scale and a 
fitting of the form 
$\omega_m=a|g- g_c|^{\alpha_{\scriptscriptstyle FL}}$ was done. The 
same was repeated for extrapolating the $t_{\perp c}$ while approaching the 
MIT from 
the Mott insulating side. This is shown in Fig.~(\ref{fig:phase_diag})(c). 
As, shown in Fig.~(\ref{fig:phase_diag})(b) and 
Fig.~(\ref{fig:phase_diag})(c), $\omega_m$ and $\Delta_g$ vanish at the 
same critical point unlike a conventional Hubbard model at $T=0$, indicating 
that the Mott MIT obtained in this model is a second order first transition.
 It is straightforward to see that each of these lines of QCPs is just a cut 
 in the surface of QCPs in the $t_\perp-U-V$ space. 
 This finding of an entire surface of Mott transitions, 
 in a `new' model, namely a bilayer KI-M model, that has not been explored 
 before, represents the main result of this work. Moreover, our $T=0$ 
 calculations show that the Mott transitions are quantum critical in nature. 
 The quantum critical surface separates two distinct phases, namely, a Kondo screened phase 
 and a Mott insulating phase.

   

   Our calculations show that the Mott quantum criticality that 
   we discuss here involves
   a complete destruction of the Fermi surface (FS) in the $f$ as well as 
   $c$-spectrum
  (discussed in detail below) and is thus not  a `large to small FS' transition,
  in contrast to the generic QCPs observed in most heavy fermion systems\cite{Si_Steglich_QCP_review}. 
  Within DMFT, the electronic 
  self energy is independent of momentum, and, hence the Mott critical points
  found here are local in nature.  
The possibility 
  of a quantum critical Mott transition has been proposed earlier 
  \cite{Senthil1, Imada_Mott_criticality}.
  Recent experiments \cite{Mott_QCP2015} on quasi-two dimensional organic 
  systems suggest the existence of a $T=0$ continuous
  Mott transition. Recent analyses of scaling behaviour of finite temperature resistivity curves 
  in the half-filled \cite{Hanna_Mott_criticality} and doped Hubbard model, computed through a continuous time quamtum Monte Carlo solver within DMFT,
  \cite{DMFT_Mott_QCP2015}
  have been interpreted as 
  evidence of hidden Mott quantum criticality.  
  The current model of coupled symmetric PAM and 
  noninteracting metal represents a new avenue where Mott quantum criticality can be studied in great detail within a 
  microscopic Hamiltonian framework. 
  Ideally one should analyse properties such as 
  the specific heat and susceptibility in the proximity 
  of a QCP and determine the critical exponents associated with them. Further the fixed points
  of the model should be found through renormalization group analysis in order
  to identify the universality class. 
We plan to carry out such analyses in future work.

\subsection{Single-particle dynamics}

  A standard paradigm for investigating the Mott transition is the 
  single band Hubbard model, where  
  the first-order MIT (at $T=0$) occurs by a continuous vanishing of $Z$ 
at a critical interaction strength with the resonance pinned at $\omega=0$ in the middle of a 
  preformed gap 
\cite{DMFT_RMP, Nozieres_JPSJ, preformed_gap1, preformed_gap2, Mott_MIT_exact}. 
However, the Mott transition is not confined 
  to just the Hubbard model but has also been observed in the symmetric PAM 
  \cite{PAM_Mott_transn1, *PAM_Mott_transn2, PAM_Mott_transn3} 
  or other heavy fermion models, albeit with a dispersing 
  $f$ band \cite{Pepin4}. 
 The Mott transition in  the symmetric  PAM with nearest neighbour 
 hybridization was justified~\cite{PAM_Mott_transn3} through the equivalence 
 of the model to
  a single-band Hubbard model close to the Mott transition 
  and
  at low frequencies ($\omega\sim ZD$).  
  The bilayer KI-M model under consideration also
  has a similar equivalence, albeit with a difference. 
  In a tiny neighbourhood of $\omega=0$, the
  $f$-density of states has the form, $D^f(\omega)=(Z^2V^2/\omega^2) D_1^c(\omega)$. Thus, from  
  (Equation~\eqref{eq:Dc}), it may be inferred that the current model is equivalent 
  to a Hubbard model with a bare bandwidth
  of $V^2/t_\perp^2$ (from the first term) at the lowest frequencies ($|\omega| \ll ZD$); while for all other frequencies, the second 
  term (gapped at the Fermi level) has a finite contribution. 
  Moreover, it is clear that the first term is a function only of 
  $\tilde{\omega}=
  \omega t_\perp^2/ZV^2$ while in the second term, a single scaling variable 
  cannot be defined.
  
  \begin{figure}[h!]
   \centerline{\includegraphics[clip=,scale=0.6]
                        {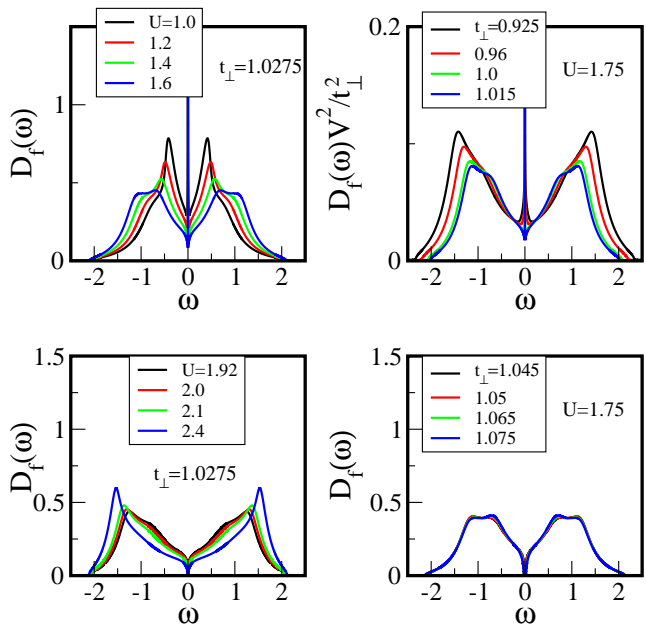}}
   \caption{The $f$ electron spectrum across a $U$ driven (left panels) 
            and a $t_\perp$ driven(right panels) QCP shown over all 
            energy scales.}
   \label{fig:spectra_all_scales}
\end{figure}

  In Fig.~(\ref{fig:spectra_all_scales}) we plot the representative 
  $f$-spectrum over all energy scales 
  for different $U$'s and $t_\perp$'s in the FL phase (top panel) and the  
  Mott insulating phase (bottom panel) at a fixed $V=0.44$. 
  In the Fermi liquid the structure on 
  all scales is similar to that of the single band HM,
  namely that there are two Hubbard bands that move away from (towards) the 
  Fermi level with increasing $U$ (increasing $t_\perp$). In the 
  Mott insulating phase at a fixed $t_\perp$ 
  (Fig.~(\ref{fig:spectra_all_scales}), left bottom panel), in accordance with the 
  conventional single band Hubbard model, $\Delta_g$ decreases with increasing 
  $U$, simultaneously pushing the Hubbard bands out. However, with increasing $t_\perp$ and a fixed $U$, the high energy Hubbard bands seem unaffected to a large extent 
   (Fig.~(\ref{fig:spectra_all_scales}), right bottom panel).

\begin{figure}[t]
    \centerline{\includegraphics[clip=,width=8cm,height=8cm]
                        {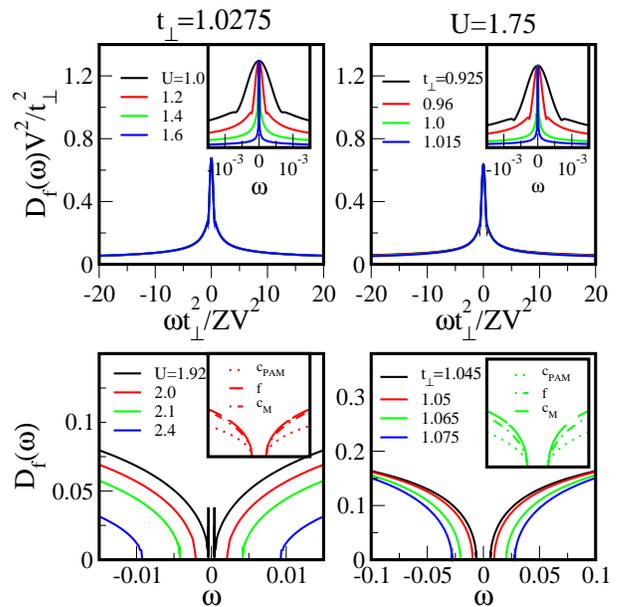}}
    \caption{\textbf{$f$-spectrum across  a $U$-driven (left panels) and
    a $t_\perp$ driven (right panels) QCP.} The top panels show the 
    collapse of the Fermi liquid
    metallic spectra when plotted as a function of 
    $\tilde{\omega}=\omega t_\perp^2/ZV^2$, while the insets show the same 
    spectra on a bare frequency scale.
    The bottom panels show the increase in the gap in the Mott insulating 
    spectra with increasing $U$ (left bottom) or increasing $t_\perp$ 
    (bottom right). The insets in the bottom panel show the $c$-spectrum of 
    the heavy fermion layer ($c_{PAM}$) and the non-interacting metallic layer 
    ($c_{M}$) on the same scale as the $f$-spectrum.}
    \label{fig:mit}
\end{figure}

  We now look into the low energy sector of the correlated ($f$) 
  and the conduction ($c$) electron spectrum of the heavy-fermion layer and 
  the non-interacting metallic layer. 
 Figure~\ref{fig:mit}
  shows the representative $f$-spectrum for the metallic case (top panels) and 
  the insulating case (bottom panels) for a  given QCP, reached either by 
  increasing $U$ (left panels) at a fixed $t_\perp=1.0275$ or by 
  increasing $t_\perp$ at a fixed $U=1.75$ for $V=0.44$. 
  The critical interaction strength for the left panels is $U_c\approx1.9$ 
  while the critical interlayer coupling for the right panels is   
  $t_{\perp c}\approx1.03$. The FL spectra, when scaled 
  by $V^2/t_\perp^2$, are seen to be universal functions of 
  $\tilde{\omega}=\omega t_\perp^2/ZV^2$ in the scaling regime 
  ($Z\rightarrow 0$ and finite ${\tilde{\omega}}$) as seen by the collapse of 
  the spectra for a range of $U$ or $t_\perp$ values (left and right panels 
  respectively). The insets in the two top panels show the same spectrum as 
  the main panels, but on a `bare' frequency scale that show the rapid 
  narrowing of the central resonance peak upon approaching the QCP. 
  The bottom panels show the Mott insulating spectra 
  that show the single-particle gap to increase when the $U$ or $t_\perp$ is 
  increased beyond the critical value. 
  The insets in the bottom panel show that the Fermi surface
  is completely destroyed since the Mott gap in the $f$-DoS
  is identical to that in either of the conduction bands. Thus it is not 
  orbital selective unlike the MIT observed in other related models 
  \cite{PAM_Mott_transn3, Pepin4}.

\begin{figure}[h!]
   \centerline{\includegraphics[clip=,scale=0.6]
                        {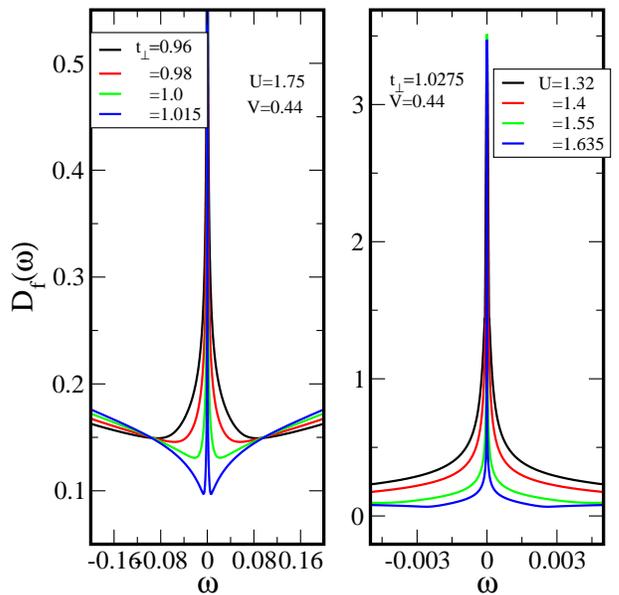}}
   \caption{The $f$ electron spectrum across a $U$ driven (left panels) and a $t_\perp$ driven (right panels) QCP (close to the transition) shown over the low  
energy scales to highlight the absence of the formation 
            of any preformed gap. The critical values are: 
$t_{\perp c}\approx1.03$ 
in the left panel and $U_c\approx1.9$ in the right panel.}
   \label{fig:no_preformed_gap}
\end{figure}

We now show in Fig.~(\ref{fig:no_preformed_gap}) that, although the overall 
spectrum of the $f$ electrons looks same as that of a single band HM, there 
lies a crucial difference: there is an absence of a preformed gap. It is now 
well known \cite{Mott_MIT_exact} that in the single band Hubbard model, the Kondo 
resonance in the metallic 
phase, resides inside a preformed insulating gap, such that at $U_{c2}$ the resonance 
vanishes leaving behind a fully formed 
insulating Mott gap \cite{preformed_gap1, preformed_gap2}.  
In fact, this serves to be a spectral fingerprint of the first-order Mott transition 
observed in such systems \cite{preformed_gap1, preformed_gap2, DMFT_RMP, 
Limelette_expt}. 
The absence of such a preformed gap in the metallic spectra of the $f$-electrons, as 
shown in Fig.~(\ref{fig:no_preformed_gap}) is crucially in agreement with the 
quantum critical nature of the Mott transition observed in the model. Here, the 
Hubbard band edges and the central Kondo resonance tails are connected by a 
finite density of states at all $t_\perp$'s 
(Fig.~(\ref{fig:no_preformed_gap}))(left panel) and at all $U$'s 
(Fig.~(\ref{fig:no_preformed_gap}))(right panel) sufficiently close to the transition, 
indicating that at the Mott transition, when the resonance vanishes, an insulating 
gap between the Hubbard bands just opens.

\section{Conclusions}
We find a surface of Mott QCPs in a bilayer KI-M model,
 which may be accessed through three athermal parameters, namely the on-site Coulomb repulsion $U$, the interlayer-hopping $t_\perp$ and the hybridization between $f$ and $c$ electrons, $V$. Experimental fabrication of precisely layered heavy fermion systems is indeed possible currently, and hence leads us to hope that the
model proposed may be realized experimentally. In fact, we find that an M-KI-M system
with the KI layer sandwiched between two non-interacting 
layers with interlayer hopping of $\sqrt{2}t_\perp$ is exactly
equivalent to the bilayer KI-M model (equation~\ref{eq:ham}), thus offering some more 
freedom in experimental fabrication. 
Additionally, we have checked that the use of a different $c$-electron DoS, 
such as a flat band or a two-dimensional tight-binding DoS with a 
van-Hove singularity at $\omega=0$ does not change the scenario of a surface 
of QCPs qualitatively. Particularly, in the context of the current prespective, 
although a tuning of $U$ and $V$ is not straightforward, the $t_\perp$ can be 
increased through uniaxial pressure, and hence the latter should be used as the experimental athermal parameter. Uniaxial pressure
  perpendicular to the planes could correspond to varying $t_\perp$, $D$ 
  and $V$ simultaneously
  in a material-specific way. Theoretically, the equation of state, 
  derived e.g. from a first-principles approach, might be able to yield the 
  possible directions of approach that correspond to a single experimental 
  variable. 
The structure of the model is not limited to a Kondo-insulator-metal 
system but similar observations are expected to occur in any three orbital 
model that can be parametrized accordingly.
However, the model Hamiltonian (equation ~\ref{eq:ham}) might not be a true representation
of the real system, since many aspects of the latter such as
$f$-orbital degeneracy, crystal field splitting, multiple conduction bands, and Hund's exchange, have been neglected. Moreover, for solving the model,
we have neglected two-dimensional dynamical correlations (due to the use of
the DMFT framework) and any symmetry breaking. 
 
Nevertheless, the KI-M system could serve as a paradigmatic theoretical model, that brings together the phenomena of Mott transition, quantum criticality,  and heavy fermions.
Our work opens up a number of theoretical directions for exploration, such
as the universality class and the relation of the QCPs 
found here to those found in impurity models~\cite{DEL_QCP_imp2014}
such as the pseudogap Anderson model \cite{LMA_pseudogap_AIM1, 
*LMA_pseudogap_AIM2} 
or the gapped Anderson model~\cite{LMA_GAIM} as well as lattice models 
incorporating competition between 
Kondo screening and Ruderman-Kittel-Kasuya-Yosida interactions~
\cite{Si_local_QCP}. 
Further investigations such as finite temperature critical scaling of transport
and thermodynamic properties of the model are in progress.

\acknowledgments 
We acknowledge financial support from CSIR, India and JNCASR.
We also acknowledge valuable discussions with M.~Jarrell, R.~C.~Budhani, 
Subroto Mukerjee and Chandrabhas Narayana. 
\bibliography{bilayer_QCP_ref}

\end{document}